\documentclass[aps,prb,10pt,twocolumn,superscriptaddress,amsmath,amssymb]{revtex4}
\usepackage{hyperref}
\usepackage{epsfig}
\usepackage{color}

\newcommand{\be}{\begin{equation}}
\newcommand{\ee}{\end{equation}}
\newcommand{\ba}{\begin{array}{rcl}}
\newcommand{\ea}{\end{array}}

\newcommand{\fe}{\mathcal{F}}

\newcommand{\emb}[1]{e^{-\beta#1}}

\newcommand{\ddp}[2]{\frac{\partial#1}{\partial#2}}
\newcommand{\de}{\mathrm{d}}
\newcommand{\ono}[1]{\frac{1}{#1}}

\newcommand{\md}[1]{\left|#1\right|}

\newcommand{\ket}[1]{\left|#1\right\rangle}
\newcommand{\avg}[1]{\left\langle#1\right\rangle}

\newcommand{\Vext}{V_{\rm ext}}

\newcommand{\Vtil}{V_{\lambda_*}}
\newcommand{\lbar}{\lambda_*}
\newcommand{\ltil}{\tilde\lambda}
\newcommand{\nep}{\mathrm{e}}
\newcommand{\Tper}{\mathrm{T}}
\newcommand{\JE}{\mathrm{JE}}
\newcommand{\rmmax}{\mathrm{max}}
\newcommand{\rmmin}{\mathrm{min}}
\newcommand{\Pleft}{\widetilde{P}}

\begin{document}
\title{Thermolubricity and the Jarzynski equality}
\author{Franco Pellegrini}
\affiliation{SISSA, Via Bonomea 265, I-34136 Trieste, Italy}
\author{Emanuele Panizon}
\affiliation{SISSA, Via Bonomea 265, I-34136 Trieste, Italy}
\author{Giuseppe E. Santoro}
\affiliation{SISSA, Via Bonomea 265, I-34136 Trieste, Italy}
\affiliation{CNR-IOM Democritos National Simulation Center, Via Bonomea 265, 
I-34136 Trieste, Italy}
\affiliation{International Centre for Theoretical Physics (ICTP), Strada Costiera 11, 
I-34151 Trieste, Italy}
\author{Erio Tosatti}
\affiliation{SISSA, Via Bonomea 265, I-34136 Trieste, Italy}
\affiliation{CNR-IOM Democritos National Simulation Center, Via Bonomea 265, 
I-34136 Trieste, Italy}
\affiliation{International Centre for Theoretical Physics (ICTP), Strada Costiera 11, 
I-34151 Trieste, Italy}

\date{\today}

\begin{abstract}

{We discuss and qualify a previously unnoticed connection between two different phenomena in the physics of nanoscale friction, general in nature and also met in experiments including sliding emulations in optical lattices, and protein force spectroscopy. 
The first is thermolubricity,  designating the condition in which a dry nanosized slider can at sufficiently high temperature and low velocity exhibit very small viscous friction $f \propto v $ despite strong corrugations that would commonly imply hard mechanical stick--slip $f \propto \log v$. 
The second, apparently unrelated phenomenon present in externally forced nanosystems, is the occurrence of negative work tails  (``free lunches'') in the work probabilty distribution, tails whose presence is necessary to fulfil the celebrated Jarzynski equality of non-equilibrium statistical mechanics. 
Here we  prove analytically and demonstrate numerically in the prototypical  classical overdamped
one-dimensional point slider (Prandtl-Tomlinson) model that the presence or absence of thermolubricity is exactly equivalent to satisfaction or violation of the Jarzynski equality.  The divide between the two regimes, satisfaction of Jarzynski with thermolubricity,  and violation of both, simply coincides with the total frictional work per cycle  falling below or above $k_BT$ respectively.  This concept can, with due caution, be extended to more complex sliders, thus inviting crosscheck experiments, such as searching  for free lunches in cold ion sliding as well as in forced protein unwinding, and 
alternatively checking for a thermolubric regime in dragged colloid monolayers.  
As an important byproduct, we derive a parameter-free formula expressing the linear velocity coefficient  of frictional dissipated power in the thermolubric viscous regime, correcting previous empirically parametrized expressions.
%
}   
\end{abstract}


\maketitle

\section{Introduction}

Understanding the physical underpinnings of sliding friction,  a centuries-old endeavour, entered a new era with the advent of nanofriction, where forced motion of a cluster, a flake, or a nanotip can be directly observed and measured under clear and physically controlled conditions. 
At that scale, most of the well-known empirical ``laws'' of macroscopic friction, familiar from high school textbooks, cease to be valid, replaced by atomistically grounded phenomenologies presently under development.~\cite{Vanossi13, Manini17} 
As long anticipated by Prandtl ~\cite{Prandtl28} the two classic modes of sliding between two solids, stick-slip or smooth (or viscous) regimes usually attained by  dry or lubricated interfaces respectively-- may actually show up {\it in the same} dry nanosystem as a function of temperature and of sliding velocity, with a crossover from stick-slip to viscous sliding realized at sufficiently high temperature and/or sufficiently low velocity.

``Thermolubricity" is the recently introduced and useful concept describing the  
viscous-like linear vanishing of dry friction with
infinitesimal sliding velocity $v\to 0$, attained  at finite temperature by the very same dry slider that will at larger velocities or lower temperatures undergo frictional stick-slip, with a $\log v$  friction dependence~\cite{Krylov2005, Jinesh2008}.
The  limiting  viscous friction is in fact a direct, universal consequence of thermal hopping over potential energy barriers. At infinitesimal sliding speed, where perfect thermal equilibrium is closest to unperturbed, diffusion establishes the Boltzmann population probability between potential minima, with consequent suppression of mechanical stick-slip among them. 
The viscous to stick-slip crossover caused by decreasing temperature and/or by increasing sliding speed is forcefully described by very extensive simulations ~\cite{Muser11} of the prototypical Prandtl-Tomlinson model --- a spring-driven single particle forced to move in a sinusoidal potential~\cite{Prandtl28, Tomlinson29}.
Experimentally, one of the freshest system where both regimes were probably observed is in the  sliding of trapped cold ion chains for which both regimes, lubric thermal sliding (essentially frictionless at very small velocity) and stick-slip (strongly frictional) were detected, 
with a neat crossover between the two~\cite{Gangloff15}.
Another example, to which we shall return
later, is probably in the velocity-dependent force-driven unfolding of a single protein~\cite{Rico13}.

In the parallel and so far disconnected arena of non-equilibrium statistical mechanics the Jarzynski equality~\cite{Jarz97, Jarz08, Crooks99} (JE) 
is  a well established exact identity stating that when a system, initially at inverse temperature $\beta= 1/(k_BT)$, is externally forced to evolve 
from a state A state to a state B characterized by a free energy difference $\Delta\fe$, the distribution of work $W$ done by the external force obeys
\be\label{JE}
\avg{\emb{W}}=\emb{\Delta\fe},
\ee
where $\avg{\ldots}$ represents the average over many realizations of the process. The physics underlying this  remarkable equation is the thermodynamically unavoidable presence of occasional rare events ---  one could call them ``free lunches'' ---  where work is gained, $W < \Delta\fe$, rather than spent, $ W >  \Delta\fe$,  as usual~\cite{Jarz08}.
However rare, free lunches are indispensable for the satisfaction of Eq.~\eqref{JE}, as one can trivially see, e.g.,  in the case  $\Delta\fe = 0$, where the RHS equals 1, while the LHS could only be smaller than 1 without the free lunches.
The Jarzynsky relation was tested in frictional simulations by Berkovich {\em et al.}~\cite{Berkovich08}, and also exploited experimentally, notably for the extraction of free energy barriers in biological nanosystems, particularly forced protein unfolding~\cite{Bustamante02}  and more recently in the dragging of colloids in optical 
lattices~\cite{Solano15},  currently employed in the emulation of friction between crystals~\cite{Bohlein12, Vanossi12}.

A common qualitative feature shared by these two seemingly unrelated phenomena and concepts, thermolubricity and Jarzynski, is that  both are verifiable in very small  systems that can thermally explore their phase space irrespective of the weak external forcing. 
Here we present a closer analysis showing now quite precisely that  for a single degree of freedom Jarzynski implies thermolubricity and viceversa, so that when the relevant sliding  parameters --- size, velocity and temperature --- are changed, the two either work or fail at exactly the same time. 
We shall do that by deriving first a parameter-free approximation to the work $W$ performed by a one-dimensional model system pulled over 
a barrier by a spring at nearly zero velocity. 
That allows a comparison of the two physical scenarios and provides an expression for viscous dissipation in the thermolubric regime which amends previous ones, now also  predicting the correct crossover from smooth sliding to stick-slip based purely on the ``static'' parameters of the system. 
The crossover is predicted to  occur when the total frictional work per cycle or event reaches the universal, parameter-independent value 
$\langle W \rangle \sim k_BT$,  a physically very satisfying result. 
These analytical results, validated by extensive one-dimensional simulations in the 1D Prandtl-Tomlinson  model are especially predictive of new experiments. Specifically, Jarzynski's free lunches should be pursued  in cold ion sliding below the critical velocity where thermolubricity was observed~\cite{Gangloff15}, and in the forced elongation of a single protein like titin where at low stretching  velocity the apparent force of extraction force became very small~\cite{Rico13}. 
Conversely, the frictional motion of a dragged colloid should be characterized as smooth and thermolubric in the regime where 
Jarzynski was found to be obeyed~\cite{Solano15}. 
In both cases, it should be possible to confirm that crossover occurs at the average work  $\langle W \rangle \sim k_B T$.
 
The present result for a single frictional contact may in future serve as a guide to understand more complex frictional situations. Among them, we shall briefly discuss the sliding of a (Frenkel-Kontorova) harmonic chain, where a single kink can play the role of the single 
sliding entity or 
contact, and the thermolubricity-Jarzynski connection neatly carries on.  
On closing, the dry friction of  genuinely multi-contact mesoscopic or macroscopic bodies  will be commented upon as a case where the one-to-one thermolubricity-Jarzynski connection is 
trivially
lost.  
At large sliding velocity, and/or  low temperature,  each contact stick-slips, Jarzynski is obviously broken, and the overall frictional work is large, 
never negative and essentially velocity-independent (Coulomb's law). 
At sufficiently low velocities and/or high temperature, on the other hand, thermolubricity may occur for both individual contacts and overall sliding,  
which then becomes viscous,
whereas Jarzynski will still be violated overall because free lunches of individual contacts will be out of phase and unobservable 
in the total work probability distribution. 

\section{Thermolubricity in the quasiadiabatic limit}

\begin{figure}[ht!]
\includegraphics[width=0.5\textwidth]{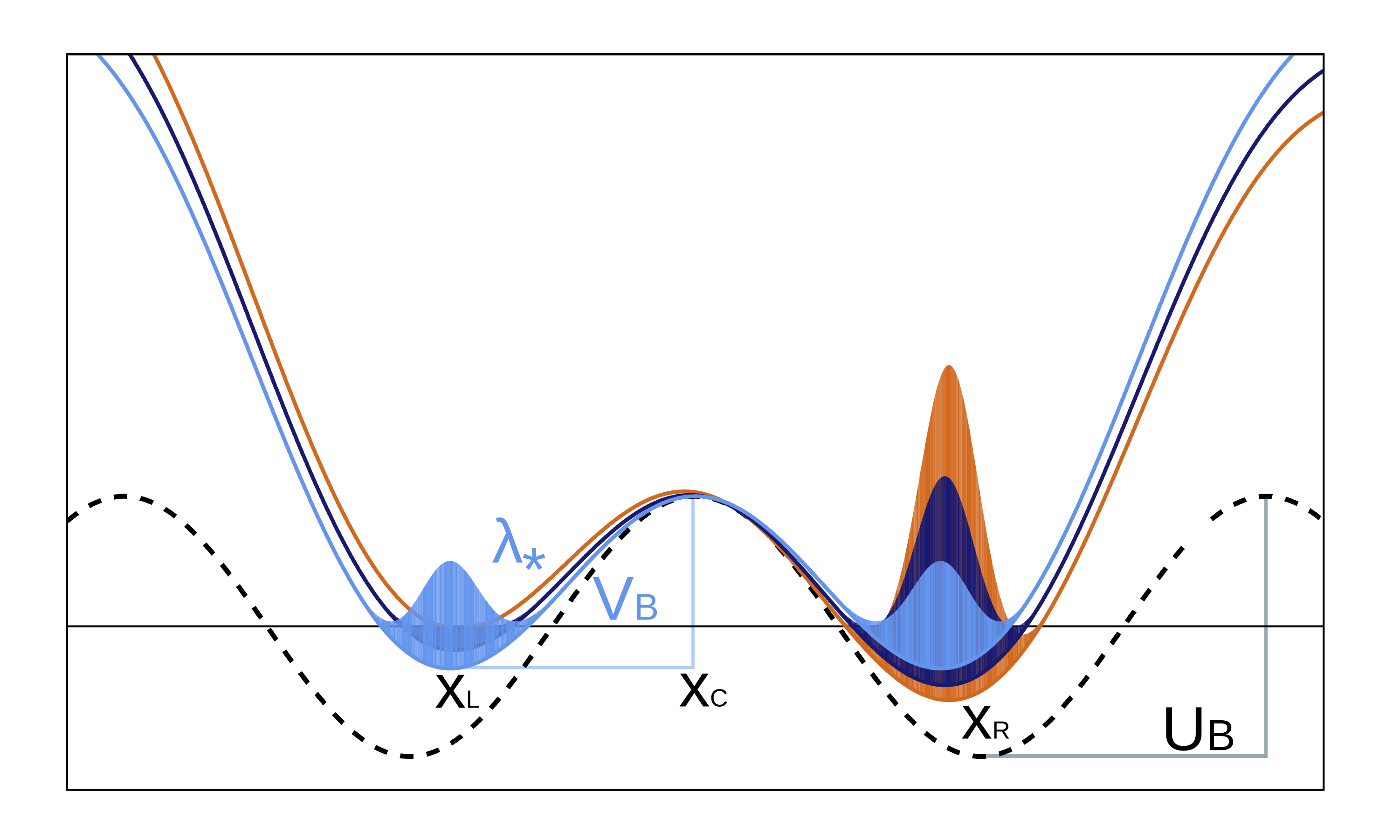}
\caption{\label{Fig1} The Prandtl-Tomlinson model potential \eqref{V_eff} (dashed line), where $U(x)\,=\,-U_0\cos(\frac{2\pi}{L}x)$. 
Continuous light-blue, blue and orange lines:  
$V_{\lambda}(x)$ for  $\lambda \,=\, 0.5,\,0.55,\,0.6$  respectively. 
For $\lambda=\lbar=0.5$, the total potential is a symmetric double well. 
Note that the effective barrier $\Delta E_{\lbar}\equiv V_B$ is different and smaller than the bare barrier $U_B=2U_0$. 
The filled curves in the right well suggest the increased equilibrium probability distributions corresponding to the three values of $\lambda$.}
\end{figure}

As announced, we start for specificity with the simplest (Prandtl-Tomlinson) model of a point particle of mass $M$ dragged in one dimension over a 
periodic sinusoidal potential $U(x)\,=\,-U_0\cos(\frac{2\pi}{L}x)$, with $U_0>0$, by a spring of stiffness $k$ moving with velocity $v$, 
so that the total potential felt by the particle is $V(x,t)=U(x)+ \frac{k}{2}(x-vt)^2$. 
If the harmonic spring potential is strong enough to single out just two lowest energy wells, left (L) and right (R), 
the elementary frictional jump takes place from one well to the next.
Before attaching the spring, the two wells, at $x_{L}=0$ and $x_{R}=L$, had equal depths $U(x_L)=U(x_R)$, and the ``bare" barrier 
between them, at $x_C=L/2$, was $U_B=U(x_C)-U(x_L) = 2U_0$.
The effective potential shape is deformed as a function of time from the bare $U(x)$ by the spring-exerted perturbation 
moving with velocity $v$. 
Calling $x_L$ and $x_R$ the coordinates of the two wells in the total effective potential (see Fig.~\ref{Fig1}), 
the global minimum switches from $x_L$ to $x_R$.
We parameterize that evolution by replacing the dragging time with a dimensionless variable $\lambda=vt/L$, where 
$\lambda$ runs from $0$ to $1$ as $t$ increase from $t=0$ to $t=\Tper=L/v$, the washboard period. 
The effective potential felt by the particle is
\be \label{V_eff}
V_{\lambda}(x)= U(x)+\frac{k}{2}\Big(x-\lambda L \Big)^2 \;.
\ee
A sketch of the evolution of the total effective potential can be seen in Fig.~\ref{Fig1}.
%

To address temperature effects at low velocity, we consider the Langevin equation for a particle in the time-dependent potential:
\be \label{eq:eqMotion_2nd}
M \ddot{x} = -\gamma \dot{x} - \nabla V_{\lambda(t)}(x(t)) + \xi(t) \;,
\ee
where $\gamma$ is a damping coefficient (also absorbing the mass $M$) and  $\xi(t)$ is a normally distributed, delta-correlated random force
with $\langle \xi(t) \rangle=0$ and  $\langle \xi(t) \xi(t') \rangle= 2 \gamma k_B T \delta(t-t')$.
We will assume the temperature $T$ to be small enough with respect to the relevant energy barrier (see below) $k_BT \ll V_B$, 
so as to avoid free diffusion between the two wells during the time $\Tper=L/v$. 
%
Physically, the slider's motion can be overdamped, as is the case in many practical situations including tip-based experiments, or underdamped, 
as for example in cold ion sliding emulations\cite{Gangloff15}, and in recent Prandtl-Tomlinson simulations.~\cite{Muser11}   
In the underdamped case, both positions and velocities are relevant, inertial terms cannot be neglected, and phase space is doubled.  In
the analytic treatment which we
develop below, we will concentrate on the (simpler) regime in which inertial forces are 
overwhelmed by a large damping (overdamped regime), 
so that
only positions play a role, and
the Langevin equation reduces to: 
\be \label{eq:eqMotion}
\dot{x}(t)=- \frac{1}{\gamma} \nabla V_{\lambda(t)}(x(t)) + \frac{1}{\gamma} \xi(t) \;.
\ee
%
In the adiabatic limit  $v\to 0$ the particle remains, thanks to the effect of the bath, infinitely close to its instantaneous 
equilibrium state at all $\lambda$, with probability distributions sketched as filled curves in Fig.~\ref{Fig1}. 
The work performed on the system by the moving spring, obtained by integrating the dragging force, 
should therefore coincide in that limit with the free energy difference between the final valley 
$R$ and the initial valley $L$, a difference which is zero in this case. 
\begin{figure}[ht!]
\includegraphics[width=0.95\columnwidth]{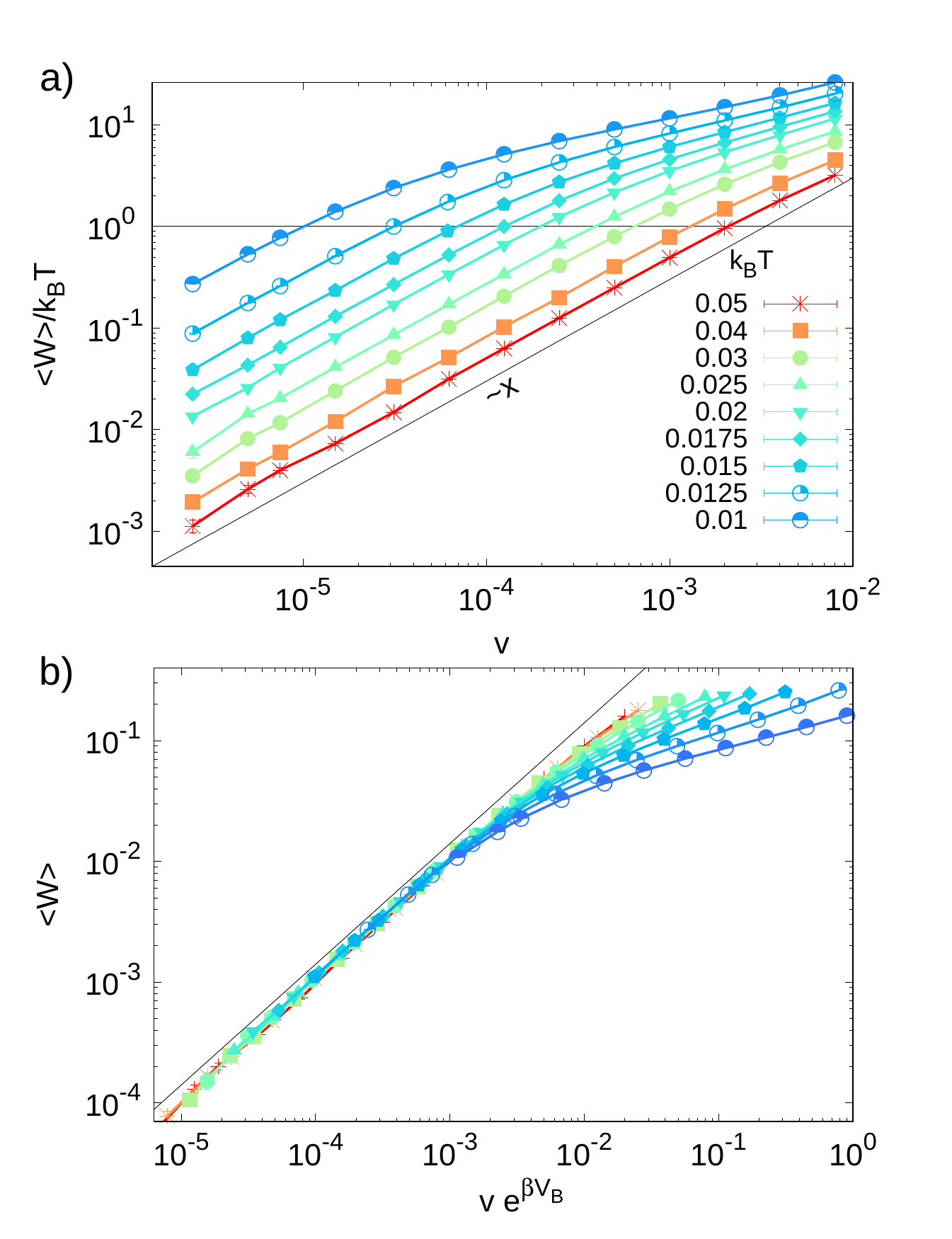}
\caption{\label{Fig2} Frictional work between consecutive minima (a distance $L$ apart) obtained with Prandtl-Tomlinson simulations. 
a) Average dissipated work --- normalized to temperature $k_BT$ --- as a function of sliding speed $v$. 
A horizontal line is drawn at $\avg{W}/k_BT\,=\,1$, separating two friction regimes, linear for $\avg{W}/k_BT<1$ and sub-linear for $\avg{W}/k_BT>1$. 
b) Same dissipated work as in (a) now presented as a function of the normalized velocity $v\nep^{\beta\Delta E_{\lbar}}\equiv v\nep^{\beta V_B}$, 
i.e., removing the temperature dependence through Eq.~\eqref{eq:Wours}. 
Model parameters $U_0=0.2$, $L=\pi$, $k=0.5$ and $\gamma=2$. 
Each point is the average of $10^5$ or $10^6$ ``slip'' events (respectively for $v<5 \times 10^{-4}$ and $v\geq 5 \times 10^{-4}$).
}
\end{figure}

In order to obtain friction, approximated by the lowest order velocity correction to the adiabatic motion, we consider the deviations of the instantaneous probability 
distribution from equilibrium. 
For that scope, we consider the Fokker-Planck (FP) equation associated with Eq.~\eqref{eq:eqMotion}, 
\begin{equation} \label{eq:FP}
\frac{\partial}{\partial t} P(x,t) = \frac{1}{\gamma} \nabla ( P \nabla V_{\lambda} ) + \frac{k_BT}{\gamma} \nabla^2 P 
\equiv \mathbb{D}_{\lambda(t)} P(x,t) \;,
\end{equation}
where $\mathbb{D}_\lambda=\frac{1}{\gamma}\left(\nabla^2V_\lambda+\nabla V_\lambda\nabla+k_BT\nabla^2\right)$
is the FP operator, and develop an adiabatic perturbation theory scheme to extract the lowest-order correction to the adiabatic motion. 
To do that, we expand the time-dependent probability distribution on the basis of the instantaneous (right) eigenvectors 
$\ket{P_i^\lambda}$ of $\mathbb{D}_\lambda$
\be
P(x,t)=P_0^{\lambda(t)}(x)+\sum_{i>0} c_i(t) P_i^{\lambda(t)}(x) \;,
\ee
with $\mathbb{D}_\lambda | P_i^\lambda\rangle = -\frac{1}{\tau_i^{\lambda}} | P_i^\lambda\rangle$.
Here $\tau_0^{\lambda}=+\infty$ corresponds to the equilibrium distribution $P_0^{\lambda}(x)$, 
and $\tau_1^{\lambda}>\tau_2^{\lambda}>\cdots$ denote the relaxation times of the higher FP eigenstates. 
Inserting this form in the FP equation \eqref{eq:FP}, the derivatives $\dot{c}_i(t)$ of the coefficients can be 
calculated to be
\be
\dot{c}_i(t)=-\frac{c_i(t)}{\tau_i^{\lambda(t)}}-\frac{1}{\Tper}
\left( \Delta_{i0}^{\lambda(t)}+\sum_{j>0} c_j(t) \Delta_{ij}^{\lambda(t)}\right) \;.
\ee
Here $\Delta_{ij}^\lambda\equiv \langle P_i^\lambda | \partial_{\lambda} P_j^\lambda\rangle$ are overlap factors defined in terms 
of the left eigenvectors $\langle P_i^\lambda|$ of the FP operator.   
For small perturbation around equilibrium, we can concentrate on the first correction only, $c_1(t)$. 
The corresponding timescale $\tau_1^{\lambda}$ --- the largest relaxation time --- should be that connected 
with the crossing of the barrier, which is the bottleneck for the systems and range of parameters we are interested in. 
As a second approximation, since we want the lowest-order correction to the adiabatic limit $\Tper\to \infty$, 
and $\dot{c}_1 = \frac{1}{\Tper} \partial_{\lambda} c_1(\lambda)$, we can safely neglect the derivative term and,
assuming $\frac{\Delta_{11}^{\lambda}}{\Tper}\ll \frac{1}{\tau_1^{\lambda}}$, finally arrive at
\be\label{cfinal}
c_1(\lambda) \simeq -\frac{1}{\Tper}\Delta_{10}^\lambda \tau_1^\lambda =  -\frac{v}{L} \Delta_{10}^\lambda \tau_1^\lambda \;,
\ee
exhibiting the expected linear dependence on velocity. 
\begin{figure}[ht!]
 \includegraphics[width=0.5\textwidth]{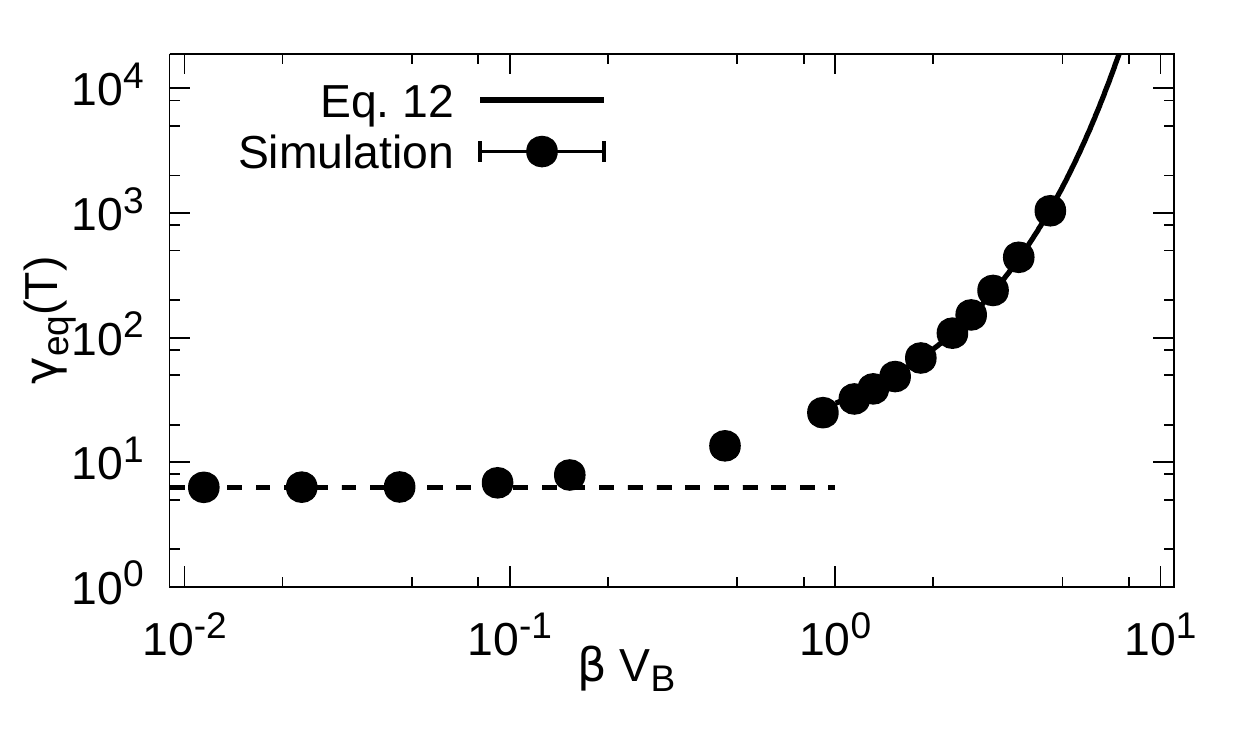}
\caption{\label{fig:comparison}
The 
linear coefficient $\gamma_{\rm eq}$ of 
frictional dissipation  $\langle W\rangle \,=\,\gamma_{\rm eq} v$.
Filled black circles,results obtained by numerical integration of the Langevin equation.
(All error bars are smaller than the size of the dots.)
In the high temperature ($\beta V_B \ll 1$) thermolubric regime 
the dissipation is low, constant and dominated by the Langevin term: $\gamma_{\rm eq}=\gamma L$ (dashed line).
%
The parameters are $U_0=0.2$, $L=\pi$, $k=0.5$ and $\gamma=2.0$. 
The solid black line shows the analytical result of Eq.~\eqref{eq:Wours} for $\beta V_B > 1$ with the fitted value of $C=0.45$. }
\end{figure}

The same linear dependence follows for the total frictional work, obtained by 
integrating the average force caused by this correction to the probability distribution 
(the equilibrium part integrates to $0$), as shown in App.~\ref{app:WPT}:
\be
\avg{W}_{\mathrm{qa}} = v\int_0^1 \! \de\lambda\, \Delta_{10}^\lambda \tau_1^\lambda
\int_{-\infty}^\infty \! \de x \; k(x-\lambda L)P_1^\lambda(x) \;.
\ee
To calculate the frictional work within our approximations, we then need to estimate the timescale $\tau_1^\lambda$ 
and the corresponding eigenstate $P_1^\lambda(x)$. 
For the first we take the inverse Kramers escape rate~\cite{Hanggi90} above the instantaneous barrier
$\Delta E_\lambda=V_\lambda(x_{\rmmax})-V_\lambda(x_{\rmmin})$:
\be
\tau_1^\lambda=\frac{2\pi\gamma \, \nep^{\beta\,\Delta E_\lambda}}{\sqrt{\md{V_\lambda''(x_{\rmmax}) V_\lambda''(x_{\rmmin})}}} \;.
\ee
We will ignore the small $\lambda$-related corrections to the position of the maximum and minima, 
so that $x_{\rmmax}=x_C$ and $x_{\rmmin}=x_{L,R}$, where we take the highest between the two inequivalent minima.
The escape time is maximum for $\lambda=\lbar$ such that the two minima are equivalent, e.g.,
$\Delta E_{\lbar}\equiv V_B =2U_0-\frac{1}{2}k(x_C-x_L)^2$, for a symmetric double well.
To approximate $P_1^\lambda(x)$, we linearly combine the equilibrium distribution 
around the two minima at each given point $\lambda$, and fix the relative 
coefficient by imposing the whole function to integrate to zero, as expected for 
a perturbation of the equilibrium probability that already integrates to one.
After some algebra, see App.~\ref{app:WPT}, we obtain for the total quasi-adiabatic work at low temperature:
\begin{equation}  \label{eq:Wours_general}
\avg{W}_{\mathrm{qa}} = C\, v\tau_1^{\lbar} \, k  \left(x_R^{\lbar}-x_L^{\lbar}\right) 
\end{equation}
where $C$ is a positive coefficient of order unity depending on the exact position of the barrier maximum with respect to the minima. 
In the symmetric case where the two minima are identical and $x_C=(x_L+x_R)/2$, we find $C\lesssim (\pi-2)/2$, 
which is very close to the fitted value $C\sim 0.43$.  
In the high temperature regime the barrier is irrelevant and free diffusion dominates. 
\begin{widetext}
Hence, denoting by $V_B\equiv\Delta E_{\lbar}$ the effective barrier, we can summarize our results as follows:
\begin{equation}  \label{eq:Wours}
\avg{W}_{\mathrm{qa}} = \left\{ 
\begin{array}{ll} 
\displaystyle C v \frac{ 2\pi \gamma \, \nep^{\beta\,V_B}}{\sqrt{\md{V_{\lbar}''(x_C)V_{\lbar}''(x_L)}}} k \left(x_R^{\lbar}-x_L^{\lbar}\right) 
& \hspace{2mm} \mbox{for} \hspace{2mm} \beta V_B \gg 1 
\vspace{2mm} \\
\gamma v L & \hspace{2mm} \mbox{for} \hspace{2mm} \beta V_B \ll 1
\end{array}
\right.
\end{equation}
Equation~\eqref{eq:Wours} is the total quasi-adiabatic work performed on the system in a frictional experiment in the thermolubric regime, 
where friction is viscous,  {\it i.e.,} linear with velocity $v$, and depends only on the geometry of the system and the 
effective damping parameter $\gamma$.
\end{widetext}
%
%
%
This parameter-free formula corrects the empirical expression originally proposed in literature for 
thermolubric friction~\cite{Krylov2005,Jinesh2008}:
\be \label{eq:Wkrylov}
\avg{W}_{\rm Ref.1} = v \frac{kL \beta U_B}{r_0} \, \nep^{\beta U_B} \;,
\ee
where $U_B = 2U_0$ is the {\it bare} barrier and $r_0$ an {\em ad-hoc} 
adjustable 
rate parameter. 
By comparison, our result 
identifies the effective friction-controlling barrier as 
the effective one
$V_{\lbar}(x)$ and not that of the bare potential $U(x)$. 
Moreover, the prefactor in our expression Eq.~\eqref{eq:Wours} is now explicitly calculated from static parameters, 
correcting Eq.~\eqref{eq:Wkrylov}, which involved the athermal rate prefactor $r_0$, sometimes used as a fitting 
parameter~\cite{Krylov2005,Muser11}.
Our derivation does not have adjustable parameters, except for the value of $C$ which 
our calculation shows, in overdamped sliding, to be $0.43$.  
Our analytical result for the average frictional work $\avg{W}$ is tested for the case of the Prandtl-Tomlinson model of Eq.~\eqref{V_eff}.
Our predicted values for the linear dependence of friction agrees extremely well with numerical simulations, as can be seen in Fig.~\ref{fig:comparison}.  We do not present here a direct comparison 
with previous Prandtl-Tomlinson simulations, performed in the underdamped regime. ~\cite{Muser11} 
However, the treatment presented here can be straightforwardly expended to that case too.   

An important outcome of our derivation is that it permits to estimate the velocity 
or temperature
where the thermolubric regime breaks down. 
The quasi-adiabatic description is only valid when for increasing velocity or decreasing temperature 
the coefficient $c_1^\lambda$ remains $\ll 1$ at all times during the evolution. 
The breakdown of the thermolubric regime will therefore take place at $\lambda=\lbar$, specifically when $c_1^{\lbar}\sim 1$:
\be\label{vmax}
v_{\rmmax}\simeq \frac{2 \, k_B T}{\tau_1^{\lbar}k(x_R^{\lbar}-x_L^{\lbar})} \;.
\ee
Insertion of this result in Eq.~\eqref{eq:Wours} yields the maximum work reached before abandoning --- upon cooling or upon speeding ---  
the thermolubric regime in favor of stick-slip: 
\be\label{Wmax}
\avg{W}_{\rmmax}^{\rm TL} \simeq 2C \, k_BT \approx \, k_B T \;.
\ee
This crossover is clearly visible in Fig.~\ref{Fig2}(a), where the normalized dissipation $\beta \langle W\rangle$ is shown as a function of 
slider velocity for a wide range of velocities. 
The curves identify clearly two regimes, linear when friction is low $\beta \langle W\rangle \lesssim 1$, sublinear when it is high.
This innocent formula, a result of this paper, shows with minimal assumptions that the transition from viscous and linear to stick-slip 
and sublinear dissipation takes place with increasing velocity or decreasing temperature when the total work is of the order of $k_BT$.
This provides a very physical, parameter-free tool to distinguish between systems in the ``proper'' thermolubric regime and other forms of lubricity. 
We should stress that this result is derived for, and applies to, a single frictional contact.  
More complex situations including multi-contact generalizations will be discussed at the end of this paper. 

\section{Jarzynski equality}

We consider now the Jarzynski equality (JE) which, as discussed in the introduction, represents an exact route to calculate variations of free energies along an 
externally forced transformation. Its importance lies in that it is valid even when the process is violent, running very far from equilibrium, as is generally the case 
in dry sliding friction between solids. 
Eq.~(1) shows that the JE involves an ensemble average which is the stumbling block 
for calculations and experiments alike.
In most experimental and numerical studies of the JE, that average is performed by repeating the transformation cycle a large number $N$ of times. 
For a given sampling 
set 
size $N$, the JE gives an approximation of the free energy difference 
\begin{equation}
\Delta F_{\JE} \sim - k_BT \log{\left( \frac{1}{N} \sum_{i=1}^N \nep^{-\beta W_i}\right)} \;,
\end{equation}
where $\{W_i\}$ is the set of dissipation works $W_i$ obtained in the different realizations of the transformation. 
It is important to note that once a protocol of 
mechanical evolution 
is fixed, e.g. by choosing the sliding velocity in our Prandtl-Tomlinson model and the temperature of the 
thermostat, the work $W$ represents a random variable following a distribution $P(W)$,
which determines the Jarzynski average.
To define that, one must focus on some important statistical features of this problem. 
For infinite sampling, i.e. in the limit $N \rightarrow \infty$, the theorem holds exactly. 
At finite $N$, i.e. for any practical purpose, the value of $\Delta_{\JE}$ will depend on $N$. 
It is therefore reasonable to study the \textit{expected} value of the Jarzynski estimator 
\begin{equation}
\Delta_{\JE}(N) = -k_B T \avg{\log \left( \frac{1}{N} \sum_{i=1}^N\nep^{-\beta W_i} \right) } \;,
\end{equation}
where $\avg{ \cdots }$ denotes the average over $N$ independent realizations, hence with probability $P(W_1) P(W_2) \cdots P(W_N)$. 
One key feature of the Jarzynski estimator is that for any $N$, the JE equality and Jensen's inequality~\cite{Gore03} 
%
imply that
$\Delta_{\JE}(N) > \Delta F$, i.e. it has a finite bias error $\Delta_{\JE}(N) - \Delta F>0$, 
a positive quantity because the region of negative dissipation $W < \Delta F$ is undersampled for any $N$. 
In our case, where $\Delta F\,=\,0$, the bias error is equal to the Jarzynski estimator $\Delta_{\JE}(N)$, and in the limit of ``worst sampling'', $N=1$, we have 
$\Delta_{\JE}(1) =  \avg{W}$. 
Verification of the JE for finite $N$ is therefore intertwined with the dependence of this bias error on the extension $N$ of sampling, which will be the topic of the following paragraphs.

\begin{figure}[t!]
\includegraphics[width=0.5\textwidth]{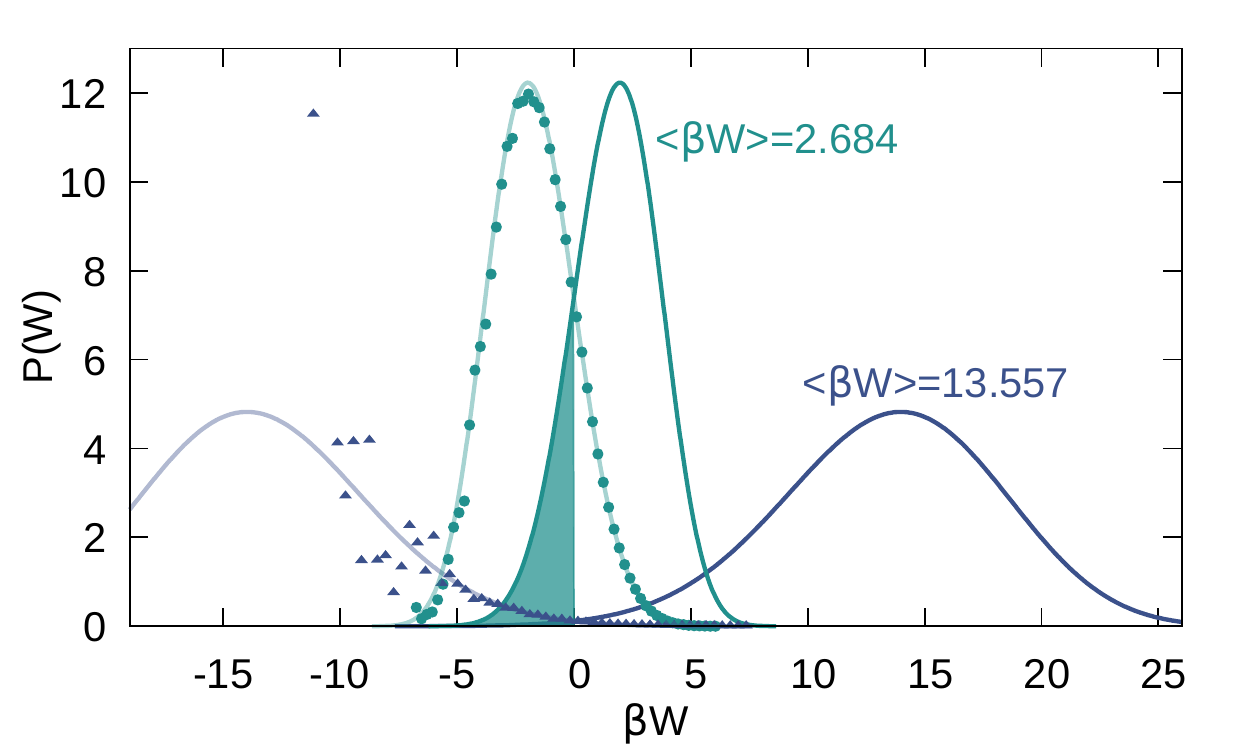}
\caption{\label{fig:JE} Probability density $P(W)$  (solid lines) for two realizations of the Prandtl-Tomlinson model 
(parameters as in panel a) of Fig.~\ref{Fig2}) for temperature $k_BT=0.0175$ and velocities $v=8 \times 10^{-3}$ and $2.5 \times 10^{-4}$, 
respectively, for dark and light red. 
The filled areas represent the total probability for a ``free lunch'', where $W < 0$. 
The shadow (mirrored) curves represent $P(-W) = P(W)\nep^{-\beta W}$, while data points indicate the numerical results 
obtained with a sample of $N=10^5$ cycles.
}
\end{figure}

We now specialize the JE, valid for a transformation process between any two states $A$ and $B$ separated by a free-energy difference $\Delta F$, 
to our case,  that is sliding friction over a periodic substrate. 
If $A$ and $B$ are two successive potential wells, the transformation takes place between two identical states: $\Delta F =0$.  
There is in addition a very helpful symmetry governing the forward process and its time-reversed one.  
One can exploit 
this symmetry by means of
Crooks' theorem~\cite{Crooks99}, stating in our case that 
\be\label{Crooks}
P(W) = P(-W) \, \nep^{\beta W} \;.
\ee
where $P(-W)$ is the time-reversed work probability distribution. 
As said, $\Delta_{\JE}(N)$ overestimates of the free energy difference because $W > \Delta F=0$ is systematically oversampled. 
Fulfilment of the JE for finite sampling depends completely on the possibility to sample those particular events where the work $W < \Delta F=0$, 
which were dubbed ``free lunches''. They are rarer by an exponential factor than their ordinary counterparts, while giving in turn an exponentially larger contribution. 
This highlights the difficulty to verify the JE both experimentally and, as in our case, numerically.  
It is known~\cite{Jarz06} however that the most relevant part of this 
negative work tail
of the distribution essential for the JE to be verified,  
and
corresponds in fact to the most common trajectories for the \textit{reverse} process. 
For a symmetric system, this represents the value of the  distribution around $P(-\avg{W})$.
In Fig.~\ref{fig:JE_test} we present two examples of the direct and inverse work probability distributions $P(W)$ and $P(-W)=P(W)\nep^{ \beta W}$ 
with sampling size $N = 10^5$ and the same sliding system.  
In both cases the free lunch contribution to the JE is centered in the neighborhood of $-\avg{W}$. 
If one restricts to $P(W)$ alone, the $N$ samples are sufficient to probe that region at small $\beta \avg{W}$ values, but they would not suffice at large ones, that is at low temperatures. 
As clarified by Fig.~\ref{fig:JE}, the Crooks formula Eq.~\eqref{Crooks} permits a perfect evaluation of that region too. 
It is also possible to estimate~\cite{Jarz06} the probability to observe the rare events where $W < 0$. 
The number $N_{\rmmin}$ of repetitions necessary to observe, in average, a single trajectory performing such work scales like
$N_{\rmmin}\sim \nep^{\beta \avg{W}}$
below which the average will be severely damaged by oversampling of the $W>0$ region. 

\begin{figure}[t!]
\includegraphics[width=0.5\textwidth]{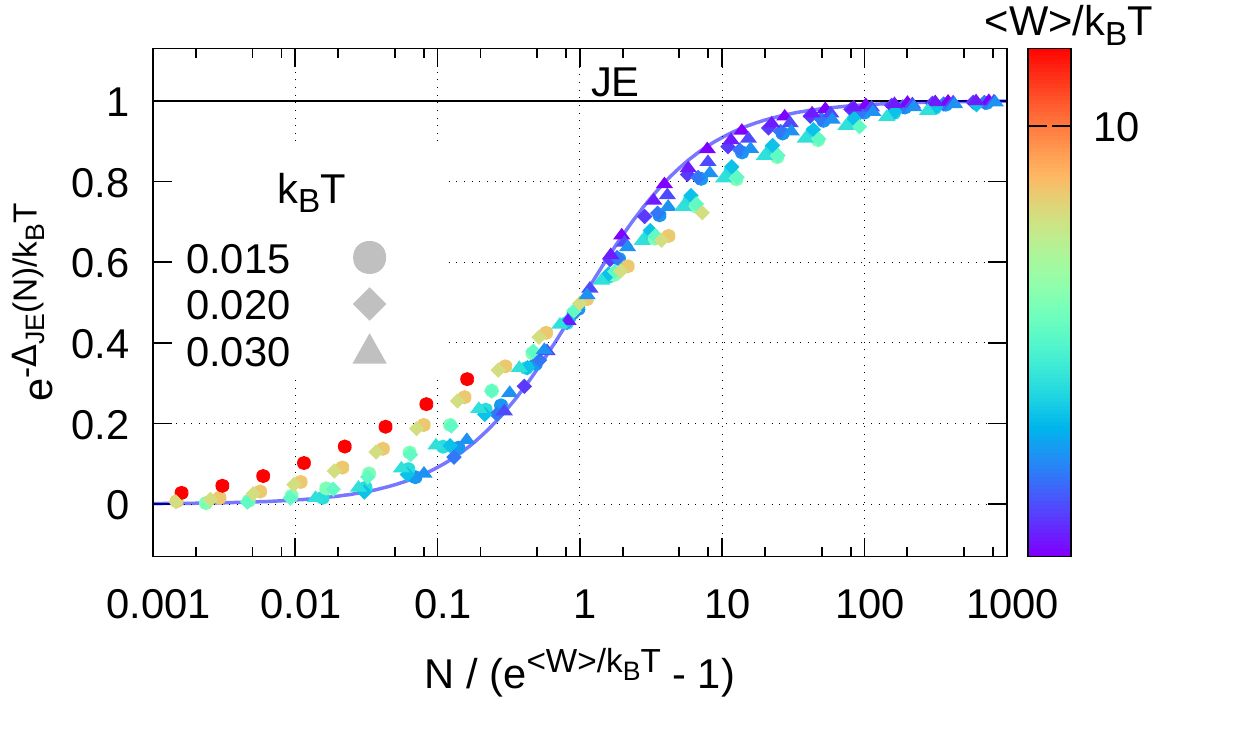}
\caption{\label{Fig4} The value of the estimated free-energy Boltzmann weight $\nep^{-\Delta_{\JE}(N)}$ 
in simulated Prandtl-Tomlinson compared to its true value $1$, i.e. to the numerical realization of the Jarzynski Equality (shown as a black line). 
The sample size $N$ is normalized by a term which is exponential in the average dissipation per cycle.
Remarkably, all curves, obtained at different temperatures and for different slider velocities, cross at the central point.
}
\end{figure}

It is revealing to study how the bias error changes as a function of the size of the sample $N$. 
In Fig.~\ref{Fig4} we show the value of $\nep^{-\beta \Delta_{\JE}(N)}$ for several cases in the Prandtl-Tomlinson system. 
We recall that the JE implies $\nep^{-\beta \Delta_{\JE}(+\infty)}=1$, and by construction $\nep^{-\beta \Delta_{\JE}(1)} = \nep^{-\beta\avg{W}}$.
Most interestingly, all curves in Fig.~\ref{Fig4} cross at 
\begin{equation}
N_{\frac{1}{2}} \equiv \nep^{\beta\avg{W}} - 1 \hspace{5mm} \mbox{where} \hspace{5mm} 
\nep^{-\beta \Delta_{\JE}(N_{\frac{1}{2}})} = \frac{1}{2} \;. 
\end{equation} 
This relation, which numerically holds in all our simulations --- for a wide range of temperatures and sliding velocities --- 
represent a striking, seemingly universal result for the bias error: $\Delta_{\JE}(\nep^{\beta \avg{W}} - 1) = k_BT \log(2)$, 
for which we could find no analytical justification, but very relevant consequences. This point acts as a precise separator between the two opposite regimes. 
For large $N > N_{\frac{1}{2}}$ all curves tend to the correct value of $1$. 
For small $\beta\avg{W}$ all the curves in Fig.~\ref{Fig4} seem to follow closely the curve $N/(N+1)$, while for larger $\beta\avg{W}$ the asymptotic true value is 
reached even more slowly.
In the opposite regime, $N_{\frac{1}{2}}$ grows \textit{exponentially} as $\beta\avg{W}$ becomes larger than unity. 
In this regime when $N \ll N_{\frac{1}{2}}$ the Jarzynski estimator is a very poor predictor of the free-energy difference.
These results show that the JE becomes exponentially hard to verify (i.e., the waiting time for free lunches become exponentially long) as soon as the average work 
performed on the system becomes of the order of a few thermal energies. 
Due to this exponential growth of the necessary sampling, we can estimate the maximum
work ``compatible'' with the experimental verification of the JE as
\be\label{WmaxJE}
\avg{W}_{\rmmax}^{\rm JE}\simeq C'\, k_BT,
\ee
with $C'$ a small constant, proportional to the logarithm of the number of events we are willing to examine.
We now see that the two crossover conditions Eq.~\eqref{Wmax} and ~\eqref{WmaxJE} coincide, proclaiming the central point of this paper:
the regime where frictional thermolubricity (linear frictional work with velocity) is realized is exactly the same where it 
should be possible to experimentally verify the JE equality without sampling an exponentially large number of trajectories. 

\section{Thermolubricity versus Jarzynski in Prandtl-Tomlinson model simulations} \label{sec:JarThermo}

To support (or falsify) the above analytical results we submitted them to direct numerical test by frictional simulations in the Prandtl-Tomlinson model, 
which is prototypical in 
friction of nanoscale systems.
In that model one performs the stochastic dynamics of a single particle dragged by a spring over a sinusoidal potential. 
Its overdamped dynamics can thus be obtained from Eq.~\eqref{eq:eqMotion}. 

We perform numerical integration of the trajectories for a wide variety of parameters and with large sampling. 
Setting $U_0=0.2$, $\gamma=2$ and $k=0.5$, we have an effective barrier $\Delta E_{\lbar}\sim 0.046$, and vary temperature and velocity in the range 
$0.01\div 0.1$ and $2.5\times 10^{-6}\div 8.0\times 10^{-3}$, respectively.
Fig.~\ref{Fig1} shows the dissipated work for $10$ different temperatures, as a function of the rescaled velocity $v\nep^{\beta V_B}$, 
following the proposed temperature dependence predicted by Eq.~\eqref{eq:Wours}.
The collapse of all curves shows how the proposed dependence is accurate for the linear regime, while different temperatures deviate at different velocities. 
Our explicit parameter-free formula \ref{eq:Wours} is shown as a black continuous line, and overestimates the observed work by 15$\%$. 
This is  still a remarkable agreement, if one considers that given the range in temperature and velocity the average friction spans over three orders of magnitude. 

To verify Eq.~\ref{Wmax}, we show in Fig.~\ref{Fig2}(a) the work rescaled over the temperature, as a function of velocity, for different temperatures, 
while a black line represents $k_BT$: it is clear how for all temperatures, the value $k_BT$ represents correctly the crossover line between linear and sublinear behaviour.

\section{Beyond the single frictional contact}

Our work so far has been to establish a physically firm and quantitative connection between thermolubricity and Jarzynski for a single frictional contact, where only one degree of freedom is connected to the external driving force. 
That outcome is in need of future consideration and generalization for extended sliding interfaces that interact through multiple contacts .
While a general and comprehensive discussion is outside of the scopes of this paper, we still anticipate here 
some
minimal extensions of the PT model to more degrees of freedom. 

\begin{figure}[ht!]
 \includegraphics[width=0.5\textwidth]{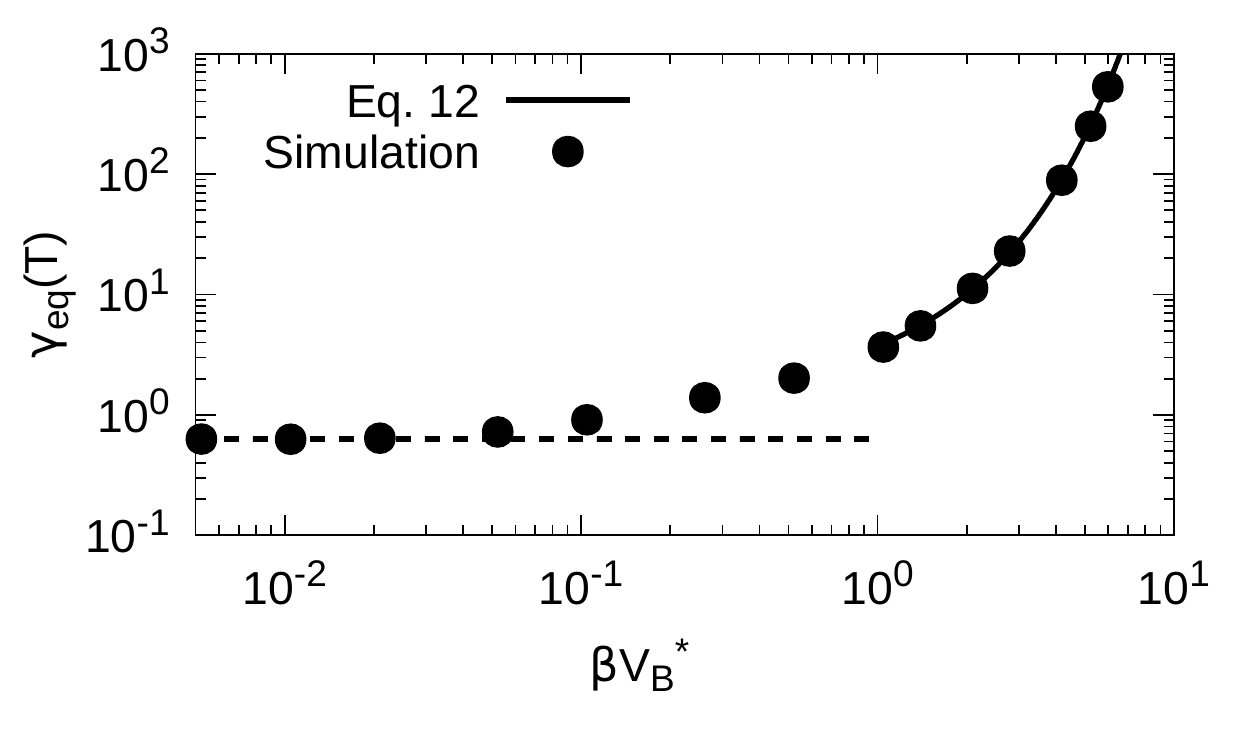}
\caption{\label{fig:kink}
Sliding in the Frenkel-Kontorova model. The linear coefficient $\gamma_{\rm eq}$ of thermolubric friction $\langle W_{single}\rangle \,=\,\gamma_{\rm eq} v$ for the 
$\frac{r_0}{a}=\frac{9}{10}$ kink. The bare parameters are $U_0=0.2$, $L=\pi$, $k=0.5$, $m=0.1$ and $\gamma=2.0$. The effective parameters are calculated as $V_B^*=0.0209$, $m^*=0.08$ and $k^*=0.4$.
(All error bars are smaller than the size of the dots.)
In the high temperature ($\beta V_B \ll 1$) regime the dissipation is constant. Dissipation is dominated by the Langevin term $\gamma_{\rm eq}=\gamma m L$ (dashed line).
The solid black line shows the analytical result of Eq.~\eqref{eq:Wours} for $\beta V_B > 1$ with the fitted value of $C=0.25$. 
The numerical results are also shown (filled black circles). }
\end{figure}

The first logical extension is to show that the frictional dissipation of a single moving kink in the sliding of an incommensurate chain of N particles in the same model periodic potential, closely resembles the sliding of the real single particle described above. In this (Frenkel-Kontorova)  model~\cite{BraunBook} the N particles interact through a nearest neighbor harmonic potential of force constant $ \kappa_{in}$ whose rest length $r_0$ differ from the periodic potential lattice parameter $a$, so that the total chain length is $L=N r_0$.
By choosing  $r_0 / a = (N-1) / N $ the chain contains a single kink, a 1D misfit dislocation. When the chain is dragged, the 
bodily
kink  motion that can be still described as a single effective degree of freedom, 
%
with position defined as $X = \sum_i^{N} x_i - C$ (where C is a constant shift term taken, e.g., such that X = 0 corresponds to the kink's position at the top of the PN potential). 
The effective potential resisting the kink motion, the Peierls-Nabarro (PN) barrier, is again periodic. The kink's effective mass ~\cite{BraunBook} $m^*$ and inverse spatial extension $\lambda^{-1}$ depend upon the original parameters $ \kappa_{in}$. The PN barrier $V_B^*$ can be made smaller and smaller, the larger and larger $ \kappa_{in}$ is relative to the effective external spring $k^*$. The effective spring constant of the driving forces on the kink, $k^*$, is defined as in the following. Each particle has a position $x_i$ and a corresponding elongation $\delta x_i$ from the driving spring attached to is $\delta x_i = x_i - x_i^0$, from which the ``elongation'' of the kink can be defined as $\delta X = \sum_i^N \delta_x$. The effective external spring constant is defined by $\frac{1}{2} k^* (\delta X)^2 = \sum_i^N \frac{1}{2}k \left(\delta x\right)^2 + A$ (where A is a constant and describes the internal strain due to the mismatch between the $a$ and $r_0$).
In this manner, the chain sliding is equivalent to that of the kink, a quasi-particle sliding in the PN potential. However when all the external springs move by a single potential lattice 
spacing 
$a$, the kink moves, 
much faster,
 for all the length $L$, through $N$ independent events, each over a PN barrier. The comparison between the dissipation of a single event $W_{single}$ and Equation ~\eqref{eq:Wours} is shown in Fig.~\ref{fig:kink}. The agreement is very good, albeit with a fitted constant C = 0.25 which differs from the case of the true single particle case. Importantly, the exponential dependence of $\gamma_{eq}$ in respect to $\beta V_{B}^*$ is recovered numerically.
We can now address 
the connection
between Jarzynski and thermolubricity for this 
model, where the Prandtl-Tomlinson sliding of a real particle is replaced by the sliding of a quasiparticle, the kink. 
For a single slip of the kink over the PN barrier, the relationship between the JE and thermolubricity is still recovered. For that case, we find once again $\avg{W_{single}} \lesssim k_BT$ as the regime boundary where both thermolubricity and JE hold.
When on the other hand one considers the total dissipation $W_{tot}=\sum_i^{N-1} \avg{W_{single}}$ over all $N-1$ 
kink slip events, 
then the Jarzynski equality is lost, since the process is equivalent to averaging over multiple slip events, and the rare negative events rapidly disappear in the average. 
Since the bath has delta-like time correlations and the motion is overdamped, all events are essentially independent.  
On the other hand, and contrary to Jarzynski's equality, thermolubricity remains valid for the total dissipation, for if single events are thermolubric, i.e. 
$\avg{W_{single}} \sim v$, then also the sum of all events will be $\avg{W_{tot}} \sim v $. 

The understanding just demonstrated, that JE and thermolubricity are one and the same thing for a single contact or degree of freedom, 
but not for a sequence of many independent events, can be naturally carried over to sliders with many degrees of freedom, 
including in general multi-contact situations, common in mesoscopic and macroscopic friction.  
Thus in the sliding of any sufficiently large or complex interface, there will be an overall thermolubric-non thermolubric transition 
as a function of temperature or of speed, but no satisfaction of the JE in either regime. 
However, the overall sliding must be imagined as the result of many individual contact motions, at least some of which poorly correlated 
with one another. If each of these uncorrelated individual contacts could hypothetically be examined, then they should 
behave as single degrees of freedom, thus obeying the JE when thermolubric.

\section{Discussion of experiments and conclusions}

The equivalence of thermolubricity and Jarzynski equality being thus discussed and validated for effectively single degrees of freedom we can 
finally examine the experimental situation.
Force-driven protein unfolding is a field where the JE 
has been
exploited, and used to extract the true equilibrium free energy cost from  non-equilibrium experiments~\cite{Bustamante02, Gore03}. 
That worked well, since at the room temperature and exceedingly low velocity conditions of these experiments, Jarzynski's relation must be 
reasonably  well obeyed.  
More recent forced-unfolding experiments ~\cite{Rico13} of titin at much larger velocities --- data also fit by formulas by Friddle {\em et al.}~\cite{Friddle12} --- 
indicate a clear change of regime around $10^2 \mu \mathrm{m/s}$ above which the JE is likely to be violated. 
On account of our results and understanding, it is highly desirable to analyse further these types of experiments with a view of establishing 
the crossover velocity, switching from viscous to stick-slip friction, and the presence/deficit of free lunches on either sides of that crossover. 

An exciting nanosystem where thermolubricity has been clearly identified is that of trapped and forced cold ions for which both thermal drift 
and stick-slip regimes are apparently accessed as a function of  velocity~\cite{Gangloff15}. 
That system too deserves now to be re-examined to detect the presence of a negative-work tails in the probability distribution, 
the satisfaction of Jarzynski in one regime but not in the other, and finally the comparison between $k_BT$ and the frictional work 
per cycle at the thermolubric/stick-slip crossover. 

Colloid layers in optical lattices have also been exploited to emulate friction~\cite{Bohlein12, Vanossi12} and their collective work distribution 
examined from Jarzynski's  point of view~\cite{Solano15}. It  should therefore be possible to extend that work by using, e.g., an optical 
tweezer to push a single colloid across the thermolubric-stick-slip crossover by monitoring both the mean frictional work and the Jarzynski tails of the distribution.

We conclude with a short list consequences that may be of direct experimental relevance and applicability:
\begin{itemize}
\item The thermolubricity regime can be directly predicted knowing only ``raw'' experimental parameters from Eq.~\eqref{vmax}. 
 This can speedup  experimental design and suggest interesting new systems where nanofriction studies can provide insightful results. 
For example, the parameters needed for Eq.~\eqref{vmax} could in a specific nanosystem be accessible from ab-initio calculations, allowing first-principle foresight into frictional thermal and velocity behaviour.

\item The work distribution in the thermolubric regime can provide experimental observations of the JE with a small number of realizations. 
In a  superlubric system the probability distribution should show large tails of negative work, and conversely the appearance negative work cycles is a telltale sign of the thermolubric regime.

\end{itemize}

\section*{ACKNOWLEDGMENTS}
This research was supported by EU FP7 under ERC-MODPHYSFRICT, Grant Agreement No. 320796.

\vspace{3mm}

\appendix

\section{The quasi-adiabatic average work in the PT model} \label{app:WPT}

What we would like to investigate is the total work $W$ performed by the external drag
on the system:
\be
W[x]=\int_0^{\Tper} \! \de t \, v \left[-\ddp{\Vext(x(t),t)}{x} \right] =
\int_0^1\! \de \lambda \, \left[ \partial_{\lambda} V_{\lambda}(x(\lambda)) \right] \;,
\ee
where we have moved the integration (and trajectory dependence) to the parameter $\lambda$.
Ideally we would like to find as much information as possible on the work distribution
$P(W)$ over multiple realizations of the dynamics. 
We will be here concerned only with the average work, which is given by
\begin{equation}
\langle W \rangle = \int_0^{1} \! \de \lambda \, \int_{-\infty}^{\infty} \! \de x \, \left[ \partial_{\lambda} V_{\lambda}(x) \right] \, P(x,\lambda) \;.
\end{equation}

We will start from the adiabatic case $v\rightarrow 0$: in this limit we can 
assume that the system has enough time to fully explore the equilibrium distribution
$P_{\rm eq}^\lambda(x)=\ono{Z_\lambda} \nep^{-\beta V_\lambda(x)}$ for each value of $\lambda$,
i.e., $P(x,\lambda)\simeq  P_{\rm eq}^\lambda(x)$.
We can then easily calculate the total work (which in this case is the same for any realization) as
\begin{eqnarray}\label{eqwork}
\avg{W}_{\mathrm{adiab}} &=& \int_0^1 \! \de \lambda \, \int_{-\infty}^{\infty} \! \de x \,  
\left[ \partial_{\lambda} V_{\lambda}(x) \right] \frac{\nep^{-\beta V_\lambda(x)}}{Z_{\lambda}} 
\nonumber \\
&=&  \int_0^1 \! \de \lambda \,  \left[ -\frac{1}{\beta} \partial_{\lambda} \ln Z_{\lambda} \right] \nonumber \\
&=& -\frac{1}{\beta}  \ln \frac{Z_1}{Z_0} = \Delta \fe \;.
\end{eqnarray}
which is just the free energy difference between the final and the initial state, as
we would expect for an adiabatic evolution. 
In the present case, the initial and final states being equivalent, we have $\Delta \fe=0$, and the 
average work vanishes in the adiabatic limit, $\avg{W}_{\mathrm{adiab}}=0$.

In the quasi-adiabatic case, we can approximate, to lowest-order in $v$,
\[ P(x,\lambda)\approx P_{\rm eq}^{\lambda}(x) + c_1(\lambda) P_1^{\lambda}(x) \;. \] 
Here, as discussed in the text, an adiabatic perturbation theory leads to:
\begin{equation}
c_1(\lambda) \approx -\frac{v}{L} \Delta_{10}^{\lambda} \tau_1^{\lambda} \;,
\end{equation}
where $\Delta_{10}^{\lambda}= \langle P_1^{\lambda} | \partial_{\lambda} P_0^{\lambda} \rangle$, 
$P_1^{\lambda}(x)$ is the first excited right eigenstate of the FP equation, and 
$\langle P_1^{\lambda}|$ the corresponding left eigenstate. 
The average work in this regime is therefore:
\begin{equation}
\avg{W}_{\mathrm{qa}} = -  v \int_0^1 \! \de \lambda \; 
\Delta_{10}^{\lambda} \tau_1^{\lambda}  F_{01}^{\lambda} \;,
\end{equation}
where we have introduced the force-like quantity:
\begin{equation}
F_{01}^{\lambda} = \frac{1}{L} \langle P_0^{\lambda} | \partial_{\lambda} V_{\lambda} | P_1^{\lambda} \rangle = 
\frac{1}{L} \int_{-\infty}^{\infty} \! \de x \, P_1^{\lambda}(x) \, \left[ \partial_{\lambda} V_{\lambda}(x) \right] \;,
\end{equation}
since the left eigenvector $\Pleft_0^{\lambda}(x)=1$.
It is here clear that the deviation from equilibrium of the average work depends
linearly on the drag velocity, since all other quantities only depend on the 
geometry of the system.
Let us now consider the quantity $\Delta_{10}^{\lambda}= \langle P_1^{\lambda} | \partial_{\lambda} P_0^{\lambda} \rangle$. 
Since
\[ \partial_{\lambda} P_0^{\lambda} = -\beta (\partial_{\lambda} V_{\lambda} ) P_0^{\lambda} - (\partial_{\lambda} \ln Z_{\lambda}) P_0^{\lambda}  \;, \]
the orthogonality $\langle P_1^{\lambda} | P_0^{\lambda}\rangle =0$ and the general fact that the left and right eigenvectors are related by
\[ \Pleft_1^{\lambda}(x) = \frac{P_1^{\lambda}(x)}{P_0^{\lambda}(x)} \;, \]
allows us to deduce that:
\begin{equation}
\Delta_{10}^{\lambda} = -\beta \langle P_1^{\lambda} | \partial_{\lambda} V_{\lambda} | P_0^{\lambda} \rangle = -\beta L F_{01}^{\lambda} \;. 
\end{equation} 
Hence the quasi-adiabatic average work is finally expressed as:
\begin{equation}
\avg{W}_{\mathrm{qa}} =  v \beta L \int_0^1 \! \de \lambda \; \tau_1^{\lambda}  \left( F_{01}^{\lambda} \right)^2  \;,
\end{equation}
which clearly shows that it is non-negative. 

As we have seen, dissipation is dominated by the dynamics where the relaxation times
are large: in a system with a barrier we will only consider the longest relaxation 
time $\tau_1$, which is related to the transition between the two minima.
Its value is maximum when the barrier is highest, i.e., when the two minima are at the same potential, 
which occurs in our case for $\lambda=\lbar=1/2$. 
We approximate the relaxation time by using the Kramer's rate formula:
\be\label{Krate}
\tau_1^{\lambda}=\frac{2\pi\gamma \, \nep^{\beta\Delta E_{\lambda}}}{\sqrt{\md{V''(x_{\rmmin})V''(x_{\rmmax})}}} \;.
\ee
Therefore, we will estimate dissipation for values of $\lambda$ around $\lbar$ such that
$V_{\lbar}(x_L)=V_{\lbar}(x_R)$, effectively moving to the variable
$\ltil=\lambda-\lbar$ and approximating:
\be
V_{\lambda}(x)\simeq \Vtil(x) - kL (x-\lbar L) \, \ltil \;.
\ee
In this regime we can estimate the barrier height as
\begin{equation}
\Delta E_{\lbar}=\Vtil(x_\rmmax)-\Vtil(x_{\rmmin})
\end{equation}
and 
\be
\Delta E_{\lambda}=\Delta E_{\lbar} -kL (x_\rmmax-x_{\rmmin}) \ltil \;.
\ee
Moreover, the energy difference between the two minima is
\be
V_{\lambda}(x_L)-V_{\lambda}(x_R) = -kL(x_L-x_R)\ltil \;.
\ee
Since the only important dependence of the relaxation time on $\ltil$ is in the barrier height, which appears in the exponential, 
we can write it as a function of its maximum value $\tau_1^{\lbar}$:
\be\label{tautmp1}
\tau_1^\lambda=\tau_1^{\lbar} \nep^{-\beta kL(x_\rmmax-x_{\rmmin})\ltil} \;.
\ee

We now need to estimate the eigenvector $P_1^{\lambda}(x)$ related to $\tau_1^{\lambda}$.  
This eigenvector must integrate to $0$ 
\[ \langle P_0^{\lambda} | P_1^{\lambda} \rangle = \int_{-\infty}^{\infty}\! \de x \, P_1^\lambda(x) = 0 \;. \]
Since $\langle P_1^{\lambda} | P_1^{\lambda} \rangle = 1$ and $\Pleft_1^{\lambda}(x) = \frac{P_1^{\lambda}(x)}{P_0^{\lambda}(x)}$, 
we also have the normalization constraint:
\be\label{Pnorm}
\int_{-\infty}^{\infty}\! \de x \, \frac{\left(P_1^\lambda(x)\right)^2}{P_0^\lambda(x)} =1 \;. 
\ee

To estimate $F_{01}^\lambda$ we need to make some assumptions on the shape of $P_1^\lambda(x)$. 
We know that for $\ltil=1/2$ this function is roughly proportional to the equilibrium distribution with opposite signs
for the left and right well. 
For $\ltil\neq 1/2$ it is reasonable to assume the same functional shape on either side, but the normalization needs to
be different to ensure the normalization constraints.
Let us break the equilibrium distribution in two functions on the left and right well:
\begin{eqnarray}
P_{L}^\lambda(x) &\equiv& P_0^\lambda(x) \, \Theta(x_C-x) \nonumber \\ 
P_{R}^\lambda(x) &\equiv& P_0^\lambda(x) \, \Theta(x-x_C)
\end{eqnarray}
where $\Theta$ is the Heaviside step function, and define the probabilities to be
in the right or left well:
\be
n_{L/R}^\lambda = \int_{-\infty}^{\infty}\! \de x \, P_{L/R}^\lambda(x) \;,
\ee
such that $n_L^\lambda+n_R^\lambda=1$ and each 
$P_{L/R}^\lambda(x)/n_{L/R}^\lambda$ is a properly normalized probability distribution.
This leads to the choice:
\begin{equation} \label{eq:P1approx}
P_1^\lambda(x) \simeq \sqrt{n_L^\lambda n_R^\lambda}
\left(\frac{P_L^\lambda(x)}{n_L^\lambda}-\frac{P_R^\lambda(x)}{n_R^\lambda} \right) \;.
\ee
If we now use the explicit form of the external potential we have
\be
F_{01}^\lambda = -k \int_{-\infty}^{\infty}\! \de x \; (x-\lambda L) \, P_1^\lambda(x) 
= -k \int_{-\infty}^{\infty}\! \de x \; x \, P_1^\lambda(x) \;. 
\ee
Plugging the approximate form of $P_1^\lambda(x)$ in Eq.~\eqref{eq:P1approx} we get:
\begin{equation}
F_{01}^\lambda \approx k \sqrt{n_L^\lambda n_R^\lambda} \left( x_R^{\lambda} - x_L^{\lambda} \right) \;,
\end{equation}
where we have introduced the averages
\begin{equation}
x_{L/R}^{\lambda} = \frac{1}{n_{L/R}^{\lambda}} \int_{-\infty}^{\infty}\! \de x \; x \, P_{L/R}^\lambda(x) \;.
\end{equation}

The last quantity we need to estimate is therefore the probability of
being in the right or left well. As a first approximation we can simply
consider this to be proportional to the depth of each well, so that
\be
n_{L/R}^\lambda \approx \frac{\nep^{-\beta V_\lambda(x_{L/R})}}
{\nep^{-\beta V_\lambda(x_L)}+\nep^{-\beta V_\lambda(x_R)}} \;.
\ee
Going back to $\ltil$ and with a little algebra we obtain
\be\label{Ftmp1}
F_{01}^\lambda \simeq\frac{k(x^\lambda_R-x^\lambda_L)}{2\cosh(\beta kL \ltil (x_R-x_L)/2)} \;,
\ee
where from now on we set $x^\lambda_{R/L}\equiv x_{R/L}^{\lbar}$. 
Considering for simplicity only the symmetric case in which $\lbar=1/2$ and $x_C=(x_R+x_L)/2$ 
we can finally plug these expressions into the average work:
\begin{widetext}
\be
\avg{W}_{\mathrm{qa}} =\frac{v\beta k^2 L \tau_1^{\lbar}(x_R-x_L)^2}{4}
\left( \int_{-\frac{1}{2}}^0 \! \de\ltil \; \frac{\nep^{ \beta kL \ltil (x_R-x_L)/2}}{\cosh^2(\beta kL \ltil (x_R-x_L)/2)}
     + \int_0^{\frac{1}{2}}  \! \de\ltil \; \frac{\nep^{-\beta kL \ltil (x_R-x_L)/2}}{\cosh^2(\beta kL\ltil (x_R-x_L)/2)} \right) \;.
\ee
\end{widetext}
Changing variable and extending the integrals to $\infty$, we can finally estimate: 
\be\label{Wtmp2}
\avg{W}_{\mathrm{qa}} \approx \frac{\pi-2}{2} vk\tau_1^{\lbar}(x_R^{\lbar}-x_L^{\lbar}) \;,
\ee
which ultimately gives us the coefficient of the linear dissipation regime.

We can now estimate the limiting value of velocity where our approximation
breaks down and the dissipation stops being linear.
A good estimate of the maximum velocity is the one giving a coefficient $c_1$ 
of order $1$ in Eq.~\eqref{cfinal}:
\be
v_{\rmmax}\simeq \rmmin_{\lambda} \frac{L}{\md{\Delta_{10}^\lambda}\tau_1^\lambda} \approx 
\frac{2 \, k_BT}{\tau_1^{\lbar}k(x_R^{\lbar}-x_L^{\lbar})} \;,
\ee
where, as we have seen, in the system we are considering the maximum time and
overlap is obtained for the first eigenstate at $\lambda=\lbar$.
Notice that $v_{\rmmax}$ is exponentially depressed by the Kramers rate $1/\tau_1^{\lbar}$
(see Eq.~\eqref{Krate}) whenever $k_BT\ll\Delta E_{\lbar}$.

This estimate for the maximum velocity can be further used with \eqref{Wtmp2} 
to compute the work at which we deviate from the linear regime: 
\be\label{Wtmp3}
\avg{W}\simeq k_BT \;.
\ee
While the prefactor could change of a factor of order unity, this result
states quite generally that for any substrate potential we will leave the 
thermolubric regime when we are competing with the bath enough to have to
supply on average more that a thermal amount of energy.

As a small aside, we can consider what would happen in the case of a system
where we can consider the force as a constant small perturbation in a fixed
potential, so that we can directly estimate dissipation in the linear response
regime from the fluctuation-dissipation theorem (FDT).
If we consider a particle performing Brownian motion in a potential, diffusion
$D$ and mobility $\mu$ are related by the FDT (or Einstein relation) $D\beta=\mu=1/\gamma$.
We can describe our friction setup by using mobility to find the force needed to 
achieve a steady state velocity $v$: $F=v/\mu=\gamma v $. 
This directly leads to a work over a length $L$ given by
\be
W=FL=\frac{vL}{\beta D} = v \gamma L \;.
\ee
clearly recovering the linear (thermolubric) regime in the high-temperature limit. 
It is sufficient to estimate the limiting velocity of the linear response regime
to be the one comparable with the natural drift velocity $D/L$ to recover the
same general result of a maximum work $W\simeq k_BT$ limiting the linear regime.




\begin{thebibliography}{30}

\bibitem{Krylov2005}
S.Yu. Krylov, K.B. Jinesh, H. Valk, M. Dienwiebel, and J.W.M. Frenken,
Phys. Rev. E {\bf 71}, 065101(R) (2005).

\bibitem{Jarz97} C. Jarzynski, Phys. Rev. Lett. {\bf 78}, 2690 (1997).

\bibitem{Jarz06} C. Jarzynski, Phys. Rev. E {\bf 73}, 046105 (2006).

\bibitem{Bustamante02} J. Liphardt, S. Dumont, S. B. Smith, I. Tinoco and C. Bustamante,
Science {\bf 296}, 5574, (2002).

\bibitem{Berkovich08} R. Berkovich, J. Klafter, and M. Urbakh, J. Phys.: Condens. Matter {\bf 20}, 354008 (2008).

\bibitem{Muser11} M. H. M\"user, Phys. Rev. B {\bf 84}, 125419 (2011).

\bibitem{Gangloff15} D. Gangloff, A. Bylinskii, I. Counts, W. Jhe, V. Vuleti\'c, Nature Physics {\bf 11}, 915 (2015). 

\bibitem{Gore03} J. Gore, F. Ritort and C. Bustamante, PNAS {\bf 100}, 12564 (2003).

\bibitem{Jinesh2008} 
K.B. Jinesh, S. Yu. Krylov, H. Valk, M. Dienwiebel, and J. W. M. Frenken, Phys. 
Rev. B { \bf 78}, 155440 (2008).

\bibitem{Jarz08} C. Jarzynski, Eur. Phys. J. B {\bf 64}, 331 (2008).

\bibitem{Vanossi13}
A. Vanossi, N. Manini, M. Urbakh, S. Zapperi and E. Tosatti, Rev. Mod. Phys. 
{\bf 85}, 529 (2013).

\bibitem{Manini17}
N. Manini,G. Mistura, G. Paolicelli, E. Tosatti, A. Vanossi, Adv. Phys. X {\bf 2}, 569 (2017).

\bibitem{Prandtl28}
L. Prandtl, Z. Angew. Math. Mech. {\bf 8}, 85 (1928).

\bibitem{Tomlinson29}  
G. A. Tomlinson, Philos. Mag. {\bf 7}, 905 (1929).

\bibitem{Rico13}
F. Rico et al.,  Science {\bf 342}, 741, (2013).

\bibitem{Friddle12} R. W. Friddle, A. Noy, J. J. De Yoreo, PNAS {\bf 109}, 13573 (2012).

\bibitem{Hanggi90}
P. H\"{a}nggi, P. Talkner and M. Borkovec, Rev. Mod. Phys. {\bf 62}, 251 (1990).

\bibitem{Crooks99}
G. E. Crooks, Phys. Rev. E {\bf 60}, 2721 (1999).

\bibitem{Bohlein12}
T. Bohlein, J. Mikhael, and C. Bechinger, Nat. Mater. {\bf 11} 126 (2012).

\bibitem{Vanossi12}
A. Vanossi, N. Manini, and E. Tosatti, PNAS {\bf 109}, 16429 (2012).

\bibitem{Solano15}
J.R. Gomez-Solano, C. July, J. Mehl, and C. Bechinger,
New J. Phys. {\bf 17}, 045026 (2015).

\bibitem{BraunBook}
O.M. Braun, Y.S. Kivshar. Springer Science \& Business Media, 2013.

\end{thebibliography}
\end{document}